\documentstyle[12pt]{article}
\def\be{\begin{equation}}
\def\ee{\end{equation}}
\def\ba{\begin{eqnarray}}
\def\ea{\end{eqnarray}}
\def\bq{\begin{quote}}
\def\eq{\end{quote}}
\def\PL{{ \it Phys. Lett.} }

\def\NP{{\it Nucl. Phys.} }
\def\PR{{\it Phys. Rev.} }

\def\gappeq{\mathrel{\rlap {\raise.5ex\hbox{$>$}}
{\lower.5ex\hbox{$\sim$}}}}

\def\lappeq{\mathrel{\rlap{\raise.5ex\hbox{$<$}}
{\lower.5ex\hbox{$\sim$}}}}
\begin{document}
\renewcommand{\theequation}{\arabic{section}.\arabic{equation}}
\newcommand{\dftwo}{(\partial\phi)^2}
\newcommand{\dffour}{(\partial\phi)^4}
\newcommand{\riemtwo}{R_{\mu\nu\sigma\rho}R^{\mu\nu\sigma\rho}}
\newcommand{\rhh}{R^{\mu\nu\sigma\rho}H_{\mu\nu\alpha}H_{\sigma\rho}{}^{\alpha}}
\newcommand{\hfour}{H_{\mu\nu\lambda}H^{\nu}{}_{\rho\alpha}H^{\rho\sigma\lambda}
H_{\sigma}{}^{\mu\alpha}}
\newcommand{\htwohtwo}{H_{\mu\rho\lambda}H_{\nu}{}^{\rho\lambda}H^{\mu\sigma\alpha}
H^{\nu}{}_{\sigma\alpha}}
\newcommand{\dP}{{\dot\Phi}}\
\newcommand{\meme}{\dot M_0 \eta \dot M_0 \eta}
\newcommand{\tr}{{\rm Tr}}
\newcommand{\dph}{{\dot\phi}}
\newcommand{\ddt}{\frac{\rm d}{{\rm d} t}}
\newcommand{\dM}{\dot M}
\newcommand{\pa}{\partial}
\newcommand{\lz}{\lambda_0}
\newcommand{\dG}{\dot G}
\newcommand{\dB}{\dot B}
\newcommand{\ch}{\rm ch}
\newcommand{\sh}{\rm sh}

\thispagestyle{empty}
\begin{flushright}
CERN-TH/96-291\\
hep-th/9610131
\end{flushright}
\vspace*{2cm}
\begin{center}
{\Large \bf  Symmetries of higher-order string gravity actions}
 \\
\vspace*{1cm}
Krzysztof A. Meissner
\footnote{Partially supported by a Polish KBN grant.}\\
\vspace*{0.2cm}
{\it CERN, 1211 Geneva 23, Switzerland}
\footnote{Permanent address: Institute for Theoretical Physics,
Ho{\.z}a 69, 00-689 Warszawa, Poland.\\
E-mail: meissner@cern.ch}\\
\vspace{3cm}
ABSTRACT
\end{center}

In this paper we explicitly prove the invariance of the 
time-dependent string gravity Lagrangian
with up to four derivatives under the global $O(d,d)$ symmetry.

\vspace*{3cm}
\begin{flushleft}
CERN-TH/96-291\\
October 1996
\end{flushleft}
\newpage

\section{Introduction}
Global, continuous symmetries
not connected with the diffeomorphism group are very rare in
gravitational systems. The first
example was discovered by Ehlers for the case of four-dimensional
pure gravity with one Killing vector. Later, it was shown
by Geroch that in the case of two Killing vectors the symmetry gets
enhanced to an infinite Kac-Moody algebra. In string theory, the
gravitational multiplet contains not only the graviton, but also a
scalar (dilaton) and the antisymmetric tensor (often referred to as
torsion). The symmetries of the Ehlers and Geroch type were also shown
in this case \cite{bakas}. Another type of symmetry in such a system
was discovered
in \cite{mv12} (without the torsion field the discrete symmetry of the
action was discovered in \cite{ven} and \cite{ts91}). It was shown
that, for the case of fields depending
only on time in an arbitrary number of dimensions (1 time, $d$ space
dimensions), the
lowest-order Lagrangian exhibits continuous, global $O(d,d)$ symmetry.
The symmetry
was later extended to the presence of matter \cite{gv} or gauge fields
\cite{hassen} and seems
to be present in a large number of string-inspired theories containing
gravity. In \cite{mv12} argument was given that the symmetry
should be present to all orders in $\alpha'$ in the $\sigma$-model
expansion (another argument was given in \cite{sen}). 
In \cite{ts91} it was argued that for the case  
without the torsion field there should be corrections to fields in the
next order in
$\alpha'$ to ensure vanishing of the $\beta$-functions and in
\cite{panvel} it was demonstrated on one specific example.
Since the inclusion of the next-order terms (like curvature
squared) can be very important for the stability of the solutions (as was
recently discussed for the case with no torsion in \cite{gmagv}), it is
the purpose of the present paper to show that the $O(d,d)$ symmetry is
explicit in the order $\alpha'$ Lagrangian of gravity coupled to the
dilaton and the antisymmetric tensor fields. There is quite a number
of authors that have calculated the higher-order effective action coming from
string amplitudes or from loop calculations in the $\sigma$-models
(see for example \cite{tm,gs}) that
sometimes do not agree with one another. 
We assumed throughout this paper that the
result of \cite{tm} is correct, and it turned out that with this
assumption the $O(d,d)$ symmetry of the quartic action can be proved.

\section{The $O(d,d)$ symmetry in the lowest order}

\setcounter{equation}{0}
In \cite{mv12} it was shown that the lowest-order string gravity
(gravity coupled to dilaton and the antisymmetric tensor) Lagrangian,
for fields depending only on cosmic time, possesses explicit
$O(d,d)$ invariance, where $d$ is the number of space dimensions. We
will recall here this construction to set the notation.
The lowest-order Lagrangian reads (throughout this paper we use the
string frame with $e^{-2\phi}$ out front, since the symmetry is
most simply realized there)
\be
\Gamma^{(0)} = \int d^{d+1}x \sqrt{-g}~ e^{-2\phi}
\left\{R+4\dftwo-\frac{1}{12}H^2\right\}.
\label{actl}
\ee
Our metric is $(-,+,\ldots,+)$ and
\be
R^{\mu}{}_{\nu\rho\sigma}=\pa_{\rho}\Gamma^{\mu}{}_{\nu\sigma}-\ldots,\
\ \ \ \ \ R_{\mu\nu}=R^{\alpha}{}_{\mu\alpha\nu},\ \ \ \ \ \ \
H_{\mu\nu\rho}=\pa_{\mu}B_{\nu\rho}+{\rm cyclic}
\ee
When fields depend only on time, it is possible to bring $g$ and $B$
to the form
\be
g_{\mu\nu}=
\left(\begin{array}{cc}
                 -1&0\\
                  0&G(t)
\end{array}\right),
\ \ \ \ \ B_{\mu\nu}=\left(\begin{array}{cc} 0&0\\ 0&B(t)\end{array}\right).
\ee

It was shown in \cite{mv12} that the action (\ref{actl}) can then be
rewritten as
\be
\Gamma^0=-\int dt e^{-\Phi}\left(\dP^2+\frac18 \tr [\meme]\right),
\label{symm0}
\ee
where
\be
\Phi=2\phi-\frac12\ln\det G,
\ee  
$\eta$ is the metric for the $O(d,d)$ group in non-diagonal form:
\be
\eta=\left(\begin{array}{cc}{\bf 0}&{\bf 1}\\
            {\bf 1}&{\bf 0}\end{array}\right)
\ee
and 
\be
M_0=\left(\begin{array}{cc}G^{-1}&-G^{-1}B\\
         BG^{-1}&G-BG^{-1}B
\end{array}\right).
\label{mzdef}
\ee
This $M_0$ has two important properties \cite{mv12}: it is symmetric and
it belongs to the $O(d,d)$ group:
\be
M_0^T=M_0,\ \ \ \ \ \ M_0\eta M_0=\eta.
\label{mprop}
\ee

The action (\ref{symm0}) is explicitly symmetric under the action of the
$O(d,d)$ group:
\be
M_0\to \Omega^T M_0\Omega,\ \ \ \ \ \ \Phi\to \Phi,
\label{symtr}
\ee
where $\Omega$ belongs to the $O(d,d)$:
\be
\Omega^T\eta\Omega=\eta.
\ee
The general $O(d,d)$ element connected to the identity can be written
as:
\be
\Omega=\exp\left(\begin{array}{cc}{A_1}&{A_2}\\
            {A_3}&{-A_1^T}\end{array}\right)\ \ \ \ A_2^T=-A_2,\ \ \
 A_3^T=-A_3.
\label{oddelem}
\ee

\section{The symmetry in the next order without torsion}
\setcounter{equation}{0}
We start with the following form of fourth order in derivatives action
in the string frame (formula 3.24 in \cite{tm})
\ba
\Gamma &=& \int d^{d+1}x \sqrt{-g}~ e^{-2\phi}
\left\{ R+4\dftwo-\frac{1}{12}H^2 \right.  \nonumber\\
&& -\alpha'\lz
\left[\riemtwo-\frac12\rhh+\right. \nonumber\\
&& \left. \left. \frac1{24}\hfour-\frac18\htwohtwo\right]
+O(\alpha'^2)\right\}
\label{ts}
\ea
($\lz=-\frac18$ for the heterotic string, $-\frac14$ for
the Bose string and 0 for the superstring).

This is the simplest possible form of the string effective action. If
one makes local redefinitions of fields, it does not change the
equations of motion (in the redefined fields); however, the
symmetry can be easily seen for one choice but impossible to guess for
another. Thus we have to try all possible redefinitions to see whether
we can bring the action to some suitable form. There are two
guidelines for the search. The first one is that the action (when
expressed in terms of time derivatives of fields and with all
integrations by parts used) contains only first derivatives of
fields. It turns out that this can always be done. The
second one is that the whole action can be written in terms of
$\Phi$ and $M_0$ defined before (but with possible corrections
of order $\alpha'$), since the symmetry is then explicit.

We start to show the techniques involved with the simpler case of
vanishing $H$ (then, of course, we do not have the full $O(d,d)$
symmetry but only some discrete subgroup); temporarily, we use
the Lagrangian:
\be
\Gamma = \int d^{d+1}x \sqrt{-g}~ e^{-2\phi}
\left\{ R+4\dftwo-\alpha'\lz\riemtwo
+O(\alpha'^2)\right\}
\label{tsnoh}
\ee

Requiring only first time
derivatives allows for the four structures of order $\alpha'$:
\be
\int d^{d+1}x\sqrt{-g}e^{-2\phi}\left[
a_1R^2_{GB}+a_2\left(R^{\mu\nu}-\frac12g^{\mu\nu}R\right)\pa_{\mu}\phi
\pa_{\nu}\phi +a_3\Box\phi\dftwo+
a_4\dffour\right]
\label{poss}
\ee
where $R_{GB}^2$ is the Gauss-Bonnet term 
\be
R_{GB}^2=\riemtwo-4R_{\mu\nu}R^{\mu\nu}+R^2
\ee
In order to transform (\ref{tsnoh}) to the form (\ref{poss}) we use the
redefinitions
\ba
\delta g_{\mu\nu}&=&\alpha'[b_1 R_{\mu\nu}+b_2\pa_{\mu}\phi
\pa_{\nu}\phi +g_{\mu\nu}(b_3R+b_4\dftwo+b_5\Box\phi)]
\nonumber\\
\delta\phi&=&\alpha'[c_1 R +c_2\dftwo+c_3\Box\phi].
\label{fred}
\ea

Under these redefinitions the action (\ref{tsnoh})
is corrected by
\ba
\delta \Gamma &=& -\int d^{d+1}x \sqrt{-g}e^{-2\phi}\left\{
\left(R^{\mu\nu}+2D^{\mu}\pa^{\nu}\phi-\frac12g^{\mu\nu}(R+4\Box\phi-
4(\pa\phi)^2)\right)\delta g_{\mu\nu}+\right. \nonumber\\
&&+\left. 2(R+4\Box\phi-
4\dftwo)\delta\phi\right\}
\label{actred}
\ea

Plugging (\ref{fred}) into (\ref{actred}) we get the
form (\ref{poss}) 
when 
\ba
\delta g_{\mu\nu}&=&-4\alpha'\lz R_{\mu\nu}\nonumber\\
\delta\phi&=&-\frac12\alpha'\lz R +2\alpha'\lz\dftwo;
\label{ftred}
\ea
the action then becomes
\be
\Gamma^{(1)}=\int \sqrt{-g}e^{-2\phi}\alpha'\lz\left[
-R^2_{GB}+16\left(R^{\mu\nu}-\frac12g^{\mu\nu}R\right)\pa_{\mu}\phi
\pa_{\nu}\phi -16\Box\phi\dftwo+16\dffour\right]
\label{actrp}
\ee

In order to write the action for fields depending only on cosmic time,
we introduce the matrix
\be
W:=G^{-1}\dot G.
\ee
Then we have
\ba
\Gamma &=& \int dt e^{-\Phi}\left\{-\dP^2+\frac14\tr W^2\right.\nonumber\\
&&\left. -\alpha'\lz\left[
\frac18\tr W^4-\frac1{16}(\tr W^2)^2+\frac13\tr
W^3\dP+\frac12(\tr W^2)\dP^2-\frac13\dP^4\right]\right\}.
\label{awnc}
\ea

It is now necessary to list all possible $O(d,d)$ invariants with
first time derivatives. There are only four of them:
\be
A_1\tr (\dM_0\eta)^4+A_2(\tr (\dM_0\eta)^2)^2+A_3\tr (\dM_0\eta)^2\dP^2+
A_4\dP^4.
\ee
Since (we still suppress the $B$-dependence!)
\be
\tr (\dM_0\eta)^4=2\tr W^4,\ \ \ \ \tr (\dM_0\eta)^2=-2\tr W^2,
\ee
we see that there is one term in the action (\ref{awnc}) that does not
belong to this class. 
In order to make the symmetry explicit we change the definition of $M$
by adding to $M_0$ a term of order $\alpha'$:
\be
M=M_0-\alpha'\lz
\left(\begin{array}{cc}-G^{-1}\dG G^{-1}\dG G^{-1}&0\\
         0&\dG G^{-1}\dG
\end{array}\right)
\label{mch}
\ee
The redefined $M$ satisfies (to order $\alpha'$) the properties
(\ref{mprop}). 

With this new definition 
the total action can be rewritten as (but still without the
antisymmetric tensor in $M$):
\ba
\Gamma &=& \int dt e^{-\Phi}\left\{-\dP^2-\frac18\tr (\dM\eta)^2\right.\nonumber\\
&&\left. -\alpha'\lz\left[
\frac1{16}\tr (\dM\eta)^4-\frac1{64}(\tr (\dM\eta)^2)^2
-\frac14(\tr (\dM\eta)^2)\dP^2-\frac13\dP^4\right]\right\}.
\label{awc}
\ea

\vspace{1cm}
\section{The full $O(d,d)$ symmetry}
\setcounter{equation}{0}
We now set to prove that (\ref{awc}) is actually the proper form of the
action after inclusion of the antisymmetric tensor.
We have to try all possible redefinitions of the action (\ref{ts}) that
give only first time derivatives. We may now use, in addition to
(\ref{ftred}), the following redefinitions:
\ba
\delta g_{\mu\nu}&=&\alpha'\lz(b_6 H^2_{\mu\nu}+b_7G_{\mu\nu}H^2)\nonumber\\
\delta\phi&=&\alpha'\lz c_4 H^2\nonumber\\
\delta B_{\mu\nu}&=&\alpha'\lz(d_1
D^{\lambda}H_{\lambda\mu\nu}+d_2H_{\lambda\mu\nu}\partial^{\lambda} \phi),
\label{fredh}
\ea
where
\be
H^2_{\mu\nu}=H_{\mu\alpha\beta}H_{\nu}{}^{\alpha\beta},\ \ \ \ {\rm and} 
\ \ \ \ \ H^2=H_{\mu\alpha\beta}H^{\mu\alpha\beta}
\ee
and the Lagrangian changes as follows:
\ba
\delta \Gamma &=& -\int d^{d+1}x \sqrt{-g}e^{-2\phi}\left\{\right.\nonumber\\
&&\left(R^{\mu\nu}+2D^{\mu}\pa^{\nu}\phi-
\frac14(H^2)^{\mu\nu}-\frac1{24}g^{\mu\nu}(12R+48\Box\phi-
48(\pa\phi)^2-H^2)\right)\delta g_{\mu\nu}+\nonumber\\
&&+\left. \frac16(12R+48\Box\phi-
48\dftwo-H^2)\delta\phi+\frac12(2\pa_{\mu}\phi H^{\mu\nu\rho}
-D_{\mu}H^{\mu\nu\rho})\delta B_{\nu\rho}\right\}.
\label{actredh}
\ea

The requirement of only first time
derivatives allows for, in addition to (\ref{poss}), the following structures:
\ba
\Gamma^{(2)}&=&\int d^{d+1}x\sqrt{-g}e^{-2\phi}\left[
a_1R^2_{GB}+a_2\left(R^{\mu\nu}-\frac12g^{\mu\nu}R\right)\partial_{\mu}\phi
\partial_{\nu}\phi +a_3\Box\phi\dftwo\right.\nonumber\\
&&+a_4\dffour
+a_5 \left(\rhh-2 R^{\mu\nu}H^2_{\mu\nu}+\frac13RH^2\right)
+a_6 H^2\dftwo\nonumber\\
&&\left. +a_7\left(D^{\mu}\partial^{\nu}\phi H^2_{\mu\nu}-\frac13\Box\phi
H^2\right)+a_8\hfour+
a_9 H^2_{\mu\nu}H^2{}^{\mu\nu}\right.\nonumber\\
&&\left. +a_{10} (H^2)^2
+a_{11}H^2{}^{\mu\nu}\pa_{\mu}\phi\pa_{\nu}\phi
+a_{12}\left(D_{\mu}H^{\mu\nu\rho}H_{\nu\rho\sigma}\pa^{\sigma}\phi+
\frac16\Box\phi H^2\right)\right].
\label{possh}
\ea
Starting from the action (\ref{ts}) and trying different redefinitions,
we finally arrive at the following form of the action:
\ba
\Gamma&=&\int \sqrt{-g}e^{-2\phi}\left\{R+4\dftwo-\frac1{12}H^2\right.
\label{finalcov}\\
&&+\alpha'\lz\left[
-R^2_{GB}+16\left(R^{\mu\nu}-\frac12g^{\mu\nu}R\right)\partial_{\mu}\phi
\partial_{\nu}\phi -16\Box\phi\dftwo+16\dffour\right.\nonumber\\
&&+\frac12\left(\rhh-2 R^{\mu\nu}H^2_{\mu\nu}+\frac13RH^2\right)
-2\left(D^{\mu}\partial^{\nu}\phi H^2_{\mu\nu}-\frac13\Box\phi
H^2\right)\nonumber\\
&&\left.\left. -\frac23 H^2\dftwo-\frac1{24}\hfour+
\frac18H^2_{\mu\nu}H^2{}^{\mu\nu}
-\frac1{144}(H^2)^2\right]\right\}.\nonumber
\ea

Making the field redefinitions of order $\alpha'$ is equivalent to
all possible substitutions of the lowest-order equations of motion in
the higher-order Lagrangian. The lowest-order equations of motion are:
\ba 
R_{\mu\nu}+2D_{\mu}\pa_{\nu}\phi-\frac14H^2_{\mu\nu}=0,&&\ \ \ \ \ 
\Box\phi-2\dftwo+\frac1{12}H^2=0\nonumber\\
R+4\Box\phi-4\dftwo-\frac1{12}H^2=0,&&\ \ \ \ \ 
D^{\lambda}H_{\lambda\mu\nu}-2H_{\mu\nu}{}^{\lambda}\pa_{\lambda}\phi=0
\label{eom}
\ea
Using them and the Bianchi identities for curvature and torsion it is
relatively straightforward to show the equivalence ``on-shell'' 
of (\ref{finalcov}) with (\ref{ts}). The action (\ref{finalcov})
corresponds to the choice in (\ref{fredh}):
\be
b_6=0,\ \ b_7=0,\ \ c_4=-\frac1{24},\ \ d_1=0,\ \ d_2=4
\label{fredhfin}
\ee

To write the result (\ref{finalcov}) for the case of fields depending
only on time, we
introduce (in addition to $W$ defined before) the matrix $Y$:
\be
Y:=G^{-1}\dot B.
\ee
We have 
\ba
\Gamma &=& \int dt e^{-\Phi}\left\{-\dP^2+\frac14\tr W^2-
\frac14\tr Y^2\right.\nonumber\\
&&+\alpha'\lz\left[-\frac18\tr W^4+\frac1{16}(\tr W^2)^2-
\frac13\tr W^3\dP
-\frac12(\tr W^2)\dP^2+\frac13\dP^4\right.\nonumber\\
&&+\frac12\tr (W^2Y^2)+\frac14\tr (WYWY)-
\frac18\tr W^2\tr Y^2+\dP \tr (WY^2)+\frac12\dP^2\tr Y^2
\nonumber\\
&&\left.\left.+\frac38\tr Y^4+\frac1{16}(\tr Y^2)^2\right]\right\}.
\label{awnctot}
\ea
In order to compare it to the $O(d,d)$ symmetric form, we need the
expressions
\ba
\tr (\dM_0\eta)^4&=& 2\tr W^4+2\tr Y^4-8\tr (W^2Y^2)
+4\tr (WYWY)\nonumber\\
\tr (\dM_0\eta)^2\dP^2&=&(-2\tr W^2+2\tr Y^2)\dP^2\nonumber\\
(\tr (\dM_0\eta)^2)^2&=&(-2\tr W^2+2\tr Y^2)^2.
\label{mexp}
\ea
We see that our result (\ref{awnctot}) contains a number of terms
that are not of this form, 
so we redefine $M$:
\be
M=M_0-\alpha'\lz
\left(\begin{array}{cc}\alpha&\beta\\
         \beta^T&\gamma
\end{array}\right),
\label{fullred}
\ee
where
\ba
\alpha&=&
-G^{-1}\dG G^{-1}\dG G^{-1}+G^{-1}\dB G^{-1}\dB G^{-1}\nonumber\\
\beta&=&
G^{-1}(\dG G^{-1}\dG-\dB G^{-1}\dB) G^{-1}B
-G^{-1}(\dG G^{-1}\dB +\dB G^{-1}\dG)\label{newmdef}\\
\gamma&=&
\dG G^{-1}\dG -\dB G^{-1}\dB-
(\dG G^{-1}\dB+\dB G^{-1}\dG) G^{-1}B\nonumber\\
&&-B(-G^{-1}\dG G^{-1}\dG G^{-1}
 +G^{-1}\dB G^{-1}\dB G^{-1})B-
B G^{-1}(\dG G^{-1}\dB+\dB G^{-1}\dG).\nonumber
\ea
To order $\alpha'$ the redefined $M$ satisfies (\ref{mprop}), so that the
redefinition is itself (time- and field-dependent) an $O(d,d)$ rotation.
To make the properties (\ref{mprop}) manifest, we write the redefinition
(\ref{fullred})  as
\be
M=\omega^T M_0 \,\omega,
\ee
where $\omega$ is in the form (\ref{oddelem}), with:
\ba
A_1&=& -\alpha'\lz\left[-\frac12\dG G^{-1}\dG G^{-1} +\frac12\dB G^{-1}\dB
G^{-1}\right]\nonumber\\
A_2&=& -\alpha'\lz\left[-\dG G^{-1}\dB  -\dB G^{-1}\dG  +\frac12(\dG G^{-1}\dG-\dB
G^{-1}\dB)G^{-1} B\right.\nonumber\\
&&\left. +\frac12 B G^{-1}(\dG G^{-1}\dG-\dB
G^{-1}\dB)\right]\nonumber\\
A_3&=&0.
\ea

With this new $M$, the action 
(\ref{awnctot}) is exactly in the form anticipated before in eq. (\ref{awc}):
\ba
\Gamma &=& \int \frac{dt}{N} e^{-\Phi}
\left\{-\dP^2-\frac18\tr (\dM\eta)^2\right.\nonumber\\
&&\left. -\frac{\alpha'\lz}{N^2}\left[
\frac1{16}\tr (\dM\eta)^4-\frac1{64}(\tr (\dM\eta)^2)^2
-\frac14(\tr (\dM\eta)^2)\dP^2-\frac13\dP^4\right]\right\}.
\label{awcfin}
\ea
We have introduced the lapse function $N$ (in the first order in
derivatives action, it
is a trivial replacement $dt\to Ndt$), since it gives one more equation
of motion (called the ``$g_{00}$'' equation in \cite{mv12}) and only
afterwards we put $N$ to 1.

This action is explicitly $O(d,d)$-invariant under (\ref{symtr}). It
looks, however, like a little miracle that the coefficients in
(\ref{ts}) coming from the string amplitudes are exactly such that
they give the
explicit $O(d,d)$ symmetry of (\ref{awcfin}). 
In comparison with the lowest-order case, now the $O(d,d)$ symmetry acts
in a more complicated way, as a rotation of not only fields but
fields with their derivatives.
 
The form of the action (\ref{actrp}) needed to exhibit the symmetry is
remarkably the same as the unique (``off-shell'') form of the action
found in \cite{mavmir} (eq. (20) there). The comparison of the full
action (\ref{finalcov}) with the result of \cite{jj} is more difficult
since there are apparent contradictions between this reference and
\cite{tm,mavmir}. However, our redefinition (\ref{fredhfin}) is
exactly the same as the redefinition used in \cite{jj} and we suspect
that the ``off-shell'' conformal
invariance also leads to the unique action (\ref{finalcov}) which is a
remarkable feature pointing out to a deeper structure behind the $O(d,d)$
symmetry.

Since the $O(d,d)$ symmetry is continuous and global, it has an
associated conserved current, which means, for a theory depending only
on time, that the current should be constant (it is an ``integrated
once'' equation of motion for $M$). In analogy to \cite{mv12}
we call this constant $A$:
\be
A={\rm const}= e^{-\Phi}\left\{M\eta\dM+2\alpha'\lz\left[
\frac12 M(\eta\dM)^3-\frac18M\eta\dM \tr
(\dM\eta)^2-M\eta\dM\dP^2\right]
\right\}
\label{consc}
\ee
where $A^T=-A$ and $A\eta M=-M\eta A$ (\cite{mv12}).

The $N$ equation reads:
\ba
0&=&-\dP^2-\frac18\tr (\dM\eta)^2\nonumber\\
&& -3\alpha'\lz\left[
\frac1{16}\tr (\dM\eta)^4-\frac1{64}(\tr (\dM\eta)^2)^2
-\frac14(\tr (\dM\eta)^2)\dP^2-\frac13\dP^4\right]
\label{neq}
\ea

Equations (\ref{consc}) and (\ref{neq}) are non-linear in fields but
(as a result of the existence of symmetry) first order in
derivatives. The analysis of these equations and their solutions will
appear in a subsequent publication \cite{gmv}.

\vspace{1cm}
{\bf Acknowledgements}

I am very grateful to Gabriele Veneziano for
earlier collaboration,
for discussions on $O(d,d)$ symmetry, and for a critical reading of the
manuscript. I wish to thank Ioannis Bakas and Maurizio Gasperini for
discussions. I also thank the organizers of
the ESI Workshop in Vienna, in August 1996, where part of this work has
been done.

\end{document}